\documentclass[twocolumn]{aastex6}
\usepackage{times}
\usepackage{ amssymb }

\usepackage{hyperref}
\usepackage{xcolor}
\usepackage{amsmath}
\definecolor{dark-red}{rgb}{0.4,0.15,0.15}
\definecolor{dark-blue}{rgb}{0.15,0.15,0.4}
\definecolor{medium-blue}{rgb}{0,0,0.5}
\hypersetup{
    colorlinks,
    linkcolor={blue},
    citecolor={blue},
    urlcolor={blue}
}
\usepackage[varg]{txfonts}

\usepackage{graphicx}
\graphicspath{{figures/}}

\usepackage{color}

\newcommand{\para}{\parallel}

\newcommand{\paren}[1]{ \left( #1 \right) }
\newcommand{\kb}{k_{\rm B}}

\renewcommand{\b}{\hat{\bb{b}}}

\newcommand{\be}{\begin{eqnarray}}
\newcommand{\en}{\end{eqnarray}}
\newcommand{\pa}{\partial}
\newcommand{\f}{\frac}

\newcommand{\mH}{m_{\rm H}}
\newcommand{\vtho}{v_{\mathrm{th}, 0}}

\newcommand{\D}[2]{\frac{\partial #1}{\partial #2}}
\newcommand\bb[1]{\mbox{\boldmath{$#1$}}}
\newcommand\bcdot{\bb{\cdot}}
\newcommand\del{\nabla}
\newcommand\btimes{\times}

\defcitealias{Ber15}{BP15}
\defcitealias{Ber16a}{BP16}

\begin{document}

\title{On Helium Mixing in Quasi-global Simulations of the Intracluster
Medium}

\author{Thomas Berlok$^1$ and Martin~ E. Pessah$^{1}$}
\affil{$^1$Niels Bohr International Academy, Niels Bohr Institute, Blegdamsvej
17, DK-2100 Copenhagen \O, Denmark;
\url{berlok@nbi.dk}
\url{mpessah@nbi.dk}}

\shorttitle{On Helium Mixing in Quasi-global Simulations of the Intracluster
Medium}
\shortauthors{Berlok \& Pessah}

\begin{abstract}
The assumption of a spatially uniform helium distribution in the intracluster
medium can lead to biases in the estimates of key cluster parameters if
composition gradients are present. The helium concentration profile in galaxy
clusters is unfortunately not directly observable. Current models addressing
the putative sedimentation are one-dimensional and parametrize the presence of
magnetic fields in a crude way, ignoring the weakly-collisional, magnetized
nature of the medium. When these effects are considered, a wide variety of
instabilities can play an important role in the plasma dynamics. In a series
of recent papers, we have developed the local, linear theory of these
instabilities and addressed their non-linear development with a modified
version of Athena. Here, we extend our study by developing a quasi-global
approach that we use to simulate the mixing of helium as induced by
generalizations of the heat-flux-driven buoyancy instability (HBI) and the
magneto-thermal instability (MTI), which feed off thermal and composition
gradients. In the inner region of the ICM, mixing can occur on few Gyrs, after
which the average magnetic field inclination angle is $\sim 30-50^{\circ}$
resulting in an averaged Spitzer parameter higher by about 20\% than the value
obtained in homogeneous simulations. In the cluster outskirts the
instabilities are rather inefficient, due to the shallow gradients. This
suggests that compositions gradients in cluster cores might be shallower than
one-dimensional models predict. More quantitative statements demand more
refined models that can incorporate the physics driving the sedimentation
process and simultaneously account for the weakly-collisional nature of the
plasma.
\end{abstract}


\section{Introduction}
\label{sec: Introduction}

The intracluster medium (ICM) of galaxy clusters is comprised of a very high
temperature and low density gas in which charged particles are bound to the
magnetic field with gyroradii that are much smaller than the mean free path
of particle collisions. In this weakly collisional medium, the weak  ($\sim \mu$G)
magnetic  field (\citealt{Car02}) channels the transport of heat and momentum,
as well as the diffusion of particles.
The anisotropic character of the weakly-collisional ICM has been found to
significantly alter its dynamical properties.
Whereas the stability of a stratified gas in the presence
of a gravitational field is governed by its entropy gradient  \citep{1947ApJ...105..305L,
1958ses..book.....S}, \cite{Bal00,Bal01}  and \cite{Qua08} found that
temperature gradients can have an important impact on the stability properties
if the plasma is weakly collisional. Two distinct instabilities were found to feed off
temperature gradients in weakly-collisional plane-parallel atmospheres, even when
their entropy increases with height.
The discovery of the magnetothermal instability (MTI, \citealt{Bal00,Bal01}),
maximally unstable when the magnetic field is perpendicular to gravity and the
temperature decreases with height and the heat-flux-driven buoyancy instability
(HBI, \citealt{Qua08}), maximally unstable when the magnetic field is parallel to
gravity and the temperature increases with height, led to a surge in research
on the stability properties of the ICM during the last decade.

These investigations considered both two and  three-dimensional simulations in local,
quasi-global, and even global settings including a variety of physical effects, for instance,
anisotropic heat conduction, Braginskii viscosity, radiative cooling and imposed turbulence \citep{Par05,Par07,Par08,Par08_MTI,Par09,Bog09,2010ApJ...712L.194P,2010ApJ...713.1332R,
2011MNRAS.413.1295M,2012MNRAS.419.3319M,Kun12,Par12,2012MNRAS.419L..29P}.
This collection of studies have led to a better understanding of a number
of fundamental issues governing
ICM plasma dynamics (see \cite{Bal16} for a recent review of the physics of the MTI and HBI).
In particular, \cite{Kun11} pointed out that  Braginskii viscosity makes the fastest
growing wavelengths for the HBI very long in the direction parallel to gravity
(see also \citealt{Gup16}). This
limited the validity of the local approaches employed thus far and
ultimately led to quasi-global studies of the HBI \citep{Lat12, Kun12}.

While the temperature distribution of the ICM is observable \citep{Vik06}, the
fact that most elements are completely ionized makes it difficult to constrain
the composition of the plasma. If present, composition gradients, as envisioned
for example by the sedimentation of helium over a Hubble time, can
lead to biases in the estimates of key cluster properties with important
implications for cosmology \citep{Mar07,Pen09}.
This has motivated the study of the long-term dynamics of heavy elements in the ICM.
As an example, the models in \cite{Pen09} predict that composition gradients can lead
to a bias of up to 20 percent in the Hubble constant if the total mass of the cluster is
estimated assuming a uniform, primordial composition  (see Figure 4 in \citealt{Pen09}).
The models for the evolution of the radial distribution of elements are one-dimensional  \citep{Fab77,Gil84,Chu03,Chu04,Pen09,Sht10} and consider the effects of magnetic
fields in rather simplified form, or ignore it altogether.

Motivated by the need of a more fundamental approach to understand the role of
magnetic fields in the dynamics of weakly-collisional media  \cite{Pes13} and \cite{Ber15}
extended the works carried out in homogeneous settings by \cite{Bal00,Bal01},
\cite{Qua08}
and \cite{Kun11} to include composition gradients. They showed that a host of
instabilities feeding off composition gradients can have an important impact on the stability
properties of the ICM. Two of these instabilities are the generalizations of the MTI and HBI,
namely the magneto-thermo-compositional instability (MTCI)
and the heat- and particle-flux-driven buoyancy instability (HPBI).
Both instabilities can be active even for isothermal atmospheres if the mean
molecular weight increases with height, even if the entropy gradient increases with height.

In order to understand how the new instabilities driven by composition gradients
saturate, \cite{Ber16a} considered the non-linear evolution of the MTCI and the HPBI
in local, isothermal settings, using a modified version of the
magnetohydrodynamics (MHD) code Athena
\citep{stone_athena:_2008}.
These simplifying assumptions made it possible to
understand some of the differences observed in the saturated state of instabilities
that are driven by either thermal or composition gradients alone. A notable
difference is that the instabilities driven exclusively by composition gradients
saturate with an average magnetic field inclination of $45^{\circ}$. This is in contrast to
the thermal instabilities where the MTI drives the magnetic field to be almost parallel to gravity
\citep{Par05} and the HBI drives the magnetic field to be almost perpendicular to gravity \citep{Par08}.

In this paper, we present the first two-dimensional (2D) quasi-global
simulations of plane-parallel atmospheres with initial equilibrium
structures inspired by the models of \cite{Pen09}, that we use to model
the inner and outer regions of the ICM. We show that the HPBI leads to
mixing of the helium content in the inner regions of the ICM and, as a
consequence, diminishes the initial gradient in composition. The inclusion
of a composition gradient leads to a $\sim20$ \% increase in heat flux to the
core at late times compared with a simulation of a homogeneous ICM.

The paper is organized as follows: In Section \ref{sec:BraMHD}, we introduce
the equations of Braginskii-MHD that we employ to model a completely ionized
plasma composed of
hydrogen and helium. In Section \ref{sec:eq_profile}, we present an equilibrium
atmosphere for the inner regions of the ICM which is based on the helium
sedimentation model of \cite{Pen09}. This atmosphere is then studied in
Section \ref{sec:lin_theory} by using a quasi-global linear theory and in
Section \ref{sec:inner_icm} by performing a suite of simulations using a
modified version of the MHD code Athena \citep{stone_athena:_2008,Ber16a}.
We also present in Section \ref{sec:outer_icm} simulations of the outer region
of the ICM, where the MTCI could be active.
We conclude in Section
\ref{sec: Discussion} by discussing the consequences of plasma instabilities
on the longterm evolution of composition gradients in the ICM as well as the
limitations of our present approach.

\section{Braginskii-MHD for a binary mixture}
\label{sec:BraMHD}

We model the intracluster medium by solving equations that evolve the
total mass density, $\rho$, momentum density, $\rho \bb{v}$, magnetic field,
$\bb{B}$, total energy density, $E$, and composition, $c$, for a weakly
collisional binary mixture of hydrogen and helium. Here, $\bb{v}$ is the
fluid velocity and the total energy density of the plasma is given by
\be
E = \f{1}{2}\rho v^2 + \f{B^2}{8\pi} + \f{P}{\gamma - 1} \ ,
\en
where $\gamma=5/3$ is the adiabatic index and $P$ is the thermal gas pressure.
We assume the magnetic field to have direction $\b =
(b_x, 0, b_z)$ and define the composition of the plasma as
\be
c \equiv \f{\rho_{\rm He}}{\rho} \ .
\en
The plasma is assumed to obey the ideal gas law
\be
P=\frac{\rho k_{\rm B} T}{\mu m_{\rm H}} \, , \label{eq:eos}
\en
where $T$ is the temperature, $\mH$ is the proton mass and
$\kb$ is Boltzmann's constant.
The mean molecular weight, $\mu$, which enters in Equation (\ref{eq:eos}), can
be shown to be related to the composition by
\be
\mu=\frac{4}{8-5c} \ , \label{eq:relation_from_c_to_mu}
\en
for a completely ionized mixture of hydrogen and helium.

The equations of motion are \citep{Pes13,Ber16a}: the continuity equation for
the total mass
\be
\f{\pa \rho}{\pa t}+\del\bcdot(\rho \bb{v}) &=& 0 \,,
\label{eq:rho}
\en
the momentum equation
\be
\D{\paren{\rho \bb{v}}}{t}
+\del \bcdot \paren{\rho \bb{vv} + P_{\mathrm{T}} \mathsf{I}
- \frac{B^2}{4\pi} \b\b}
&=&
-\del \bcdot \Pi
+ \rho \bb{g}  , \,
\en
the energy equation
\be
\D{E}{t} + \del \bcdot \left[ \paren{E+P_{\mathrm{T}}}\bb{v} -
\frac{\bb{B}\paren{\bb{B\cdot v}}}{4\pi}\right]
&=& -\del \bcdot \bb{Q}_{\rm s}
- \del \bcdot
\paren{\Pi \bcdot \bb{v}} +\rho \bb{g\cdot v} \ ,
\label{eq:E}  \nonumber \\
\en
the induction equation for the magnetic field
\be
\f{\pa \bb{B}}{\pa t}&=&\del\btimes(\bb{v}\btimes\bb{B}) \,,
\label{eq:b}
\en
and the continuity equation for the helium mass
\be
\D{\paren{c\rho}}{t}+\del\bcdot(c \rho \bb{v}) &=&-\del\bcdot\bb{Q}_{\rm c}\,.
\label{eq:c}
\en
In these equations, $P_{\mathrm{T}}$ is the total pressure (gas + magnetic)
and $\bb{g} = (0, 0, -g)$ is the gravitational acceleration which we assume to
be constant.

The equations include terms that take into account the influence of three
anisotropic effects. These anisotropic effects arise because the plasma is
weakly collisional and weakly magnetised. In this regime the charged particles
are effectively bound to the magnetic field and collisions occur
primarily along the magnetic field. This makes the assosicated transport
phenomena be directed along the magnetic field.

Electrons,
which are responsible for heat conduction, can in this way create a heat flux
along the magnetic field given by
\be
\bb{Q}_{s}=-\chi_\para\hat{\bb{b}}\hat{\bb{b}}\cdot\nabla T,
\label{eq:heat-flux}
\en
where $\chi_\para$ is the Spitzer heat conductivity \citep{1962pfig.book.....S}.

Similarly, the continuity equation for the helium density includes a flux of
composition along the magnetic field given by \citep{1990ApJ...360..267B,Pes13}
\be
\bb{Q}_{c}=-D\hat{\bb{b}}\hat{\bb{b}}\cdot\nabla c \, ,
\label{eq:helium-flux}
\en
where $D$ is the diffusion coefficient.

Finally, conservation of the first adiabatic invariant of the ions can lead
to anisotropy in the pressure tensor with
differences in the parallel ($p_\para$) and perpendicular ($p_\perp$)
pressures.
This pressure difference results in gradients in velocity-components along the
magnetic field being viscously damped. This effect, called Braginskii
viscosity \citep{Bra}, is described by the viscosity tensor
\be
\Pi = - 3\rho \nu_\para \paren{\b\b -\f{1}{3} \mathbf{I} } \paren{ \b\b -
\frac{1}{3}\mathbf{I} }  \bb{:} \del \bb{v}  \ ,
\en
where $\nu_\para$ is the viscosity coefficient and $\mathbf{I}$ is a unit
tensor.

Expressions for the dependence of $\chi_\para$, $D$ and $\nu_\para$ for an
ionized mixture of
hydrogen and helium can be found in the Appendix of \cite{Ber15}. More details
about the utility of Braginskii MHD and its
range of applicability can be found in
\citet{schekochihin_plasma_2005,Kun12,Pes13} and references therein.

For future reference we also define the plasma-$\beta
=  8\pi P/B^2 = 2{v_{\rm th}^{2}}/{v_{\rm A}^{2}}, $ where $v_{\rm A}=
{B}/\sqrt{4 \pi \rho}$ is the Alfv{\'e}n speed and $v_{\rm th} =
\sqrt{P/\rho}$ is the thermal speed.


\section{Equilibrium profile}
\label{sec:eq_profile}

In order to understand the quasi-global linear dynamics arising from Equations
(\ref{eq:rho}-\ref{eq:c}), we derive an equilibrium profile for a model plane parallel atmosphere,
which has proven to be useful for capturing key aspects of the plasma dynamics in galaxy clusters.
We assume that gravity can be modelled locally via a constant acceleration, $g$, and
that the magnetic field is purely vertical, i.e., $b_x = 0$ and $b_z=1$. The magnetic
field is assumed to be weak enough that we can neglect its contribution to the total pressure
gradient responsible for hydrostatic equilibrium. Nevertheless, the vertical weak
magnetic field can enable a background heat flux in the vertical direction. Therefore,
in order for the  model atmosphere to be in equilibrium we  require
$\nabla \cdot \bb{Q}_{c} =
0$ and $\nabla
\cdot \bb{Q}_{c} = 0$.
In what follows, we ignore helium diffusion and assume $D = 0$ such that the second condition is
trivially satisfied.

Motivated by the models considered in \cite{Pen09},
which result in helium concentration profiles that peak off-center,
we consider a situation where the composition of the plasma increases
linearly outwards from the bottom of the atmosphere, i.e., center of the cluster,
as
\be
c\left(z\right)=c_{0}+ s_{\rm c} z \ ,
\en
where $s_{\rm c} = (c_Z-c_0)/L_Z$ is the slope in composition. Notice that the
mass concentration of helium, $c(z)$, is related to the mean molecular weight,
$\mu(z)$, by
\be
\mu(z)=\frac{4}{8-5c(z)} \ ,
\en
for a completely ionized plasma of hydrogen and helium.
The equilibrium needs to fulfill
\be
\frac{d}{dz}\left(\chi_\para \frac{dT}{dz}\right)  =  0 \ ,
\label{eq:heat_diffusion_equilbrium}
\en
where $\chi_\para$ is a function of temperature $T(z)$ and composition,
$c(z)$. We will for simplicity assume that $\chi_\para$ depends only on
temperature
as $\chi_\para = \chi_{\para, 0} (T/T_0)^{5/2}$, i.e. we assume that the dependence on
composition can be neglected\footnote{For the profiles employed here, the
dependence on composition is much weaker than the dependence on temperature
\citep{Pes13}. The maximum error incurred by using this approximation is
less than 5 \% on the value of $\chi_\para$ at the top of the atmosphere.}.
With
this assumption, Equation
(\ref{eq:heat_diffusion_equilbrium}) is identical to the one derived in
\cite{Lat12} and it decouples from the rest of the equations yielding the
solution
\be
T(z)=T_{0}\left(1+\zeta z\right)^{2/7} \ ,
\en
where $\zeta L_Z = \left( T_Z / T_0 \right) ^{7/2} -1 $ and $T_Z$ $\left( T_0
\right) $ is the temperature at the top (bottom) of the atmosphere. Using the
equation of state
\be
P(z) = \frac{\rho(z)\,k_{\rm B} T(z)}{\mu(z)\,m_{\rm H}} \, ,
\en
we
solve the equation for hydrostatic equilibrium
\be
\D{P}{z}  =  -\rho g \ ,
\en
and find
\be
P(z) = P_0 e^{h(0)-h(z)} \ ,
\en
where the function $h(z)$ is related to a Gauss hypergeometric function,
${}_{2}F_{1}$, as
\be
h(z) = \f{28(1+\zeta z)^{5/7} }{5 H_0 \alpha} \,\,
{}_{2}F_{1}\left(\f{5}{7},1;\f{12}{7};
\f{5s_{\rm c}\mu_0}{\alpha} (1+\zeta z) \right) . \ \: \;
\en
In deriving the above result we have introduced the scale height at the bottom
of the atmosphere
\be
H_0 = \f{\kb T_0}{\mu_0 \mH g} \ ,
\en
and the parameter
\be
\alpha = 5 s_{\rm c} \mu_0 + 4 \zeta \ .
\en
The density, $\rho(z)$, can then be found from Equation (\ref{eq:eos}).

The values of the constants used for this equilibrium atmosphere are
inspired
by the model of \cite{Pen09}. In physical units, the model atmosphere has
height $L_Z = 2H_0 = 80$ kpc,
corresponding to the region between $r/r_{500} = 0.01$ and $r/r_{500} = 0.06$
with $r_{500} = 1.63$ Mpc in their model.  The values for the
temperature and composition of the plasma at the top and bottom of the
atmosphere are given by $T_Z = 9.6$ keV, $T_0=5.8$ keV, $c_Z=0.62$, and
$c_0=0.52$, respectively.
\begin{figure}[t]
\centering
\includegraphics{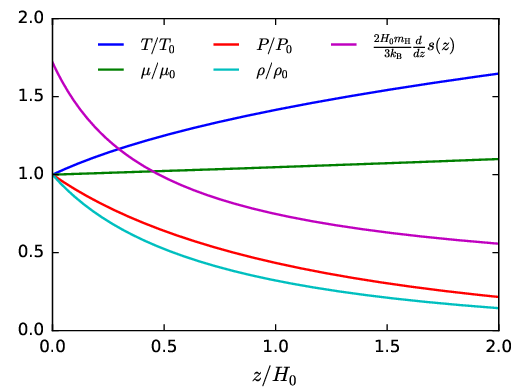}
\caption{Equilibrium atmosphere inspired by the sedimentation model of
\cite{Pen09} and the radial temperature profile of \cite{Vik06}. The
temperature (blue) and the mean molecular weight (green) increase with radius
while pressure (red) and density (magenta) decrease with radius at this radial
distance in the cluster model. The derivative of the entropy (purple) is
positive, indicating stability according to Equation (\ref{eq:schwarzschild}).
}
\label{fig:equilibrium}
\end{figure}
In Figure \ref{fig:equilibrium} we show $P(z)/P_0$, $\rho(z)/\rho_0$, $T
(z)/T_0$, $\mu(z)/\mu_0$ and $(2H_0\mH/3\kb)\,(ds/dz)$, where the entropy per
unit mass is defined by
\be
s = \f{3\kb}{2 m_{\rm{H}}} \ln \left[\left(\frac{P}{P_0}\right)
\left(\frac{\rho}{\rho_0}\right)^{-5/3}\right] \ .
\en
Note that, even though the composition increases with height, the Ledoux criterion
for stability \citep{1947ApJ...105..305L} (which is the generalization of the Schwarzschild
criterion, \citealt{1958ses..book.....S}, valid  for a heterogeneous collisional medium) is not
violated. Indeed,
for this model atmosphere the entropy is an increasing function of height, i.e.,
\be
\f{ds}{dz} > 0 \, \label{eq:schwarzschild} \,,
\en
as illustrated in Figure \ref{fig:equilibrium}. As we will discuss next, the
instabilities present in the linear theory and in the simulations are due
to the weakly collisional nature of the plasma.

\section{Quasi-global linear theory}
\label{sec:lin_theory}

The HPBI has its fastest growth rates on radial scales longer than a scale
height when Braginskii viscosity is included in the analysis \citep{Ber15}.
The local linear mode analysis is therefore not strictly valid as it assumes
that the radial scales are much shorter than a scale height. This problem has
previously been found for the HBI in \cite{Kun11} and solved
by introducing a quasi-global linear theory for the HBI in \cite{Lat12}.
In the same vein, in this section, we develop a quasi-global linear theory for the HPBI
by considering a model atmosphere with a non-uniform mean molecular weight.

The purpose of deriving a quasi-global theory is to predict the growth rates
of the instability as a function of perpendicular wavenumber, $k_x$, in order
to understand whether the instability will grow on astrophysically relevant
timescales. Furthermore, the eigenmodes obtained from linear theory can
also be used to compare with the linear stage of simulations using Athena.
Such a comparison will serve as a test of our modified version of Athena in
Section \ref{subsec:Athena}.

\subsection{Equations of Motion and Relevant Parameters}
\label{subsec:equation_of_motion}

The equations governing the quasi-global, linear dynamics for the perturbations
are obtained from the equations of Braginskii-MHD for a binary mixture \citep{Pes13,Ber15}
by using a Fourier transform along the $x$-coordinate but retaining the
$z$-derivatives
explicitly in order to relax the local approximation.
Therefore, the perturbations are  calculated in terms of the complex Fourier
coefficients, $\tilde{f}(k_x,z)$, which we assume to have a time dependence, $\exp(\sigma
t)$, where $\sigma$ is the (in general complex) eigenvalue. We assume that
$\chi_\para = \chi_{\para, 0} (T/T_0)^{5/2}$ and $\nu_\para = \nu_{\para, 0}
\,\rho_0/\rho\,(T/T_0)^{5/2}$, i.e.
$\chi_\para$ and $\nu_\para$ only depend on the composition through the
constants
$\chi_{\para,0}$ and $\nu_{\para,0}$. We also introduce a flux function such
that $\bb{B} = \del \btimes (A \bb{\hat{y}})$. The initial condition is $A = B
x$ which is equivalent to $\bb{B} = B \bb{\hat{z}}$. In the present work, we
only consider the case of $D = 0$. In this case, and with the above caveats,
the quasi-global linearized equations are: the continuity equation
\be
\sigma\frac{\delta\rho}{\rho} &=& -ik_x\delta v_{x}-\left(\frac{d\ln\rho}{dz}
+\frac{\partial}{\partial z}\right)\delta v_{z},
\label{eq:first_linearized}
\en
the $x$-component of the momentum equation
\be
\sigma\delta v_{x} &= &-ik_x\frac{T}{\mu}\left(\frac{\delta\rho}{\rho}
+\frac{\delta T}{T}-\frac{\delta\mu}{\mu}\right)+\frac{2}{\rho\beta_{0}}
\left(k_x^{2}-\frac{\partial^{2}}{\partial z^{2}}\right)\delta A
 \nonumber \\
&&- \frac{ik_x T^{5/2} }{3 \rho \, \rm Re_0}\left(2\frac{\partial\delta v_{z}}
{\partial z}-ik_x\delta v_{x}\right),
\en
the $z$-component of the momentum equation
\be
\sigma\delta v_{z} &=&-\frac{T}{\mu}\left(\frac{\delta T}{T}-
\frac{\delta\mu}{\mu}\right)\frac{d\ln P}{dz}-
\frac{T}{\mu}\frac{\partial}{\partial z}
\left(\frac{\delta\rho}{\rho}+\frac{\delta T}{T}-
\frac{\delta\mu}{\mu}\right) \nonumber \\
&& + \frac{2 T^{5/2}}{3\rho \rm Re_0}
\left(\frac{5}{2}\frac{d\ln T}{dz}+
\frac{\partial}{\partial z}\right)\left(2\frac{\partial\delta v_{z}}
{\partial z}-ik_x\delta v_{x}\right), \quad
\en
the entropy equation (with $\gamma = 5/3$)
\be
\frac{3}{2}\sigma\frac{\delta T}{T}  &=&  -ik_x\delta v_{x}-
\left(\frac{3}{2}\frac{d\ln T}{dz}+\frac{\partial}{\partial z}\right)
\delta v_{z}+ \nonumber  \\
&&\frac{1}{P \,\rm Pe_0}\left(\frac{\partial^{2}}{\partial z^{2}}
\left(T^{7/2}\frac{\delta T}{T}\right)+q_{0}ik_x
\frac{\partial\delta A}{\partial z}\right) \ ,
\en
the induction equation
\be
\sigma\delta A=-\delta v_{x} \ ,
\en
and the equation for the mean molecular weight, $\mu$,
\be
\sigma\frac{\delta\mu}{\mu} &=& - \delta v_{z}\frac{d\ln \mu}{dz}.
\label{eq:last_linearized}
\en
Here, the perturbation to the flux function is related to the
perturbation to the magnetic field by $\delta \bb{B} = \del
\btimes (\delta A\, \bb{\hat{y}})$.

Equations (\ref{eq:first_linearized}-\ref{eq:last_linearized}) have been
written in dimensionless variables by scaling $\mu$ with $\mu_0$, $T$ with
$T_0$, the velocities with $\vtho$, $\sigma$ with $\sigma_0=
t_0^{-1}=\vtho/H_0$, $\delta A$ with
$B_0 H_0$ and $z$ with $H_0$ such that $k_x$ is scaled with $1/H_0$. The
background heat flux parameter $q_0$, is given by
\be
q_0 = -T^{7/2} \f{d\ln T}{dz} \ ,
\en
in dimensionless variables.
At the bottom of the atmosphere, $z = 0$, the Peclet number is given by
\be
\textrm{Pe}_0 = \f{v_{\mathrm{th},0}\,P_0\,H_0}{\chi_{\para, 0}\,T_0}
\approx 70 \ ,
\en
and the Reynolds number is given by
\be
\textrm{Re}_0 = \f{v_{\mathrm{th},0}\,\rho_0\,H_0 }{\nu_{\para, 0}}
\approx 3800 \ .
\en
These parameters were found by using Equations (B6) and (B7) in \cite{Ber15}
to estimate the
values of $\chi_{\para, 0}$ and $\nu_{\para, 0}$. Following \cite{Lat12} and
\cite{Kun12} we furthermore take the plasma-$\beta$ at the bottom of the
atmosphere to be
\be
\beta_0 = 10^{5} \ .
\en

\subsection{Solutions obtained with  a pseudo-spectral method}
\label{subsec:Pseudo}

The linearized quasi-global equations (Equations
\ref{eq:first_linearized}-\ref{eq:last_linearized}) are solved using a pseudo-
spectral method, in a manner similar to the analysis presented in
\cite{Lat12}, see also \cite{Boyd}. We discretize the six equations on a
Chebyshev-Gauss-Lobatto roots grid transformed onto the domain $z = [0,2]$.
The grid has $N$ grid points, where $N = 200$, and the resulting algebraic
equations constitute a generalized eigenvalue problem of size $6N$. We use the
same boundary conditions as \cite{Lat12} which means that $\delta v_z$,
$\delta T/T$, and $\pa _z \delta A$ are set to zero at the boundaries. The
latter condition corresponds to $\delta B_x = 0$ at the boundaries, which implies
that the field remains vertical there. We furthermore impose $\delta \mu /\mu
= 0$ at the boundaries.

\begin{figure}[t]
\centering
\includegraphics{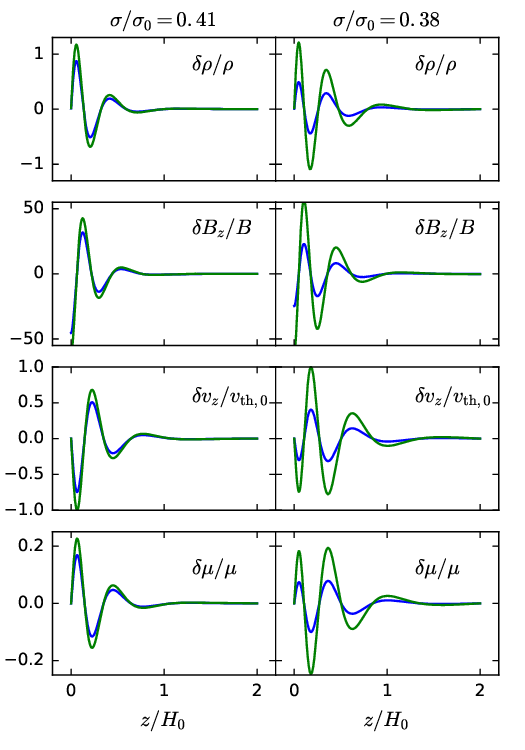}
\caption{Select components of the eigenmode for the fastest (left, $m=1$) and
second fastest (right, $m=2$) eigenmode with $n = 5$ and the real (imaginary)
part of the eigenmode shown in blue (green). A general rule seems to be that
the fastest growing modes have a small vertical extent while slower growing
modes have a larger vertical extent. This trend also appears in the
homegeneous setting \citep{Lat12} and it is consistent with the simulations
presented in Section \ref{sec:inner_icm}.}
\label{fig:quasi_global}
\end{figure}

Equations (\ref{eq:first_linearized})-(\ref{eq:last_linearized}) only depend
on the value of $k_x H_0 = 2 \pi n$ where $n$ is the horizontal mode number.
All other parameters are set by our model. For each value of $k_x$, there are a
number of modes which we designate with the vertical mode number $m$,
where $m= 1$ is the fastest growing mode, $m = 2$ labels
the second fastest growing mode, and so on.
We show the solution for $n = 5$ in Figure \ref{fig:quasi_global}, where
the left panel shows the $m = 1$ mode which is confined to the lower
region of the domain. The right panel shows the $m = 2$ mode which has a
slower growth rate but a larger vertical extent than the $m=1$ mode. This is a
general property of the solutions: The vertical extent of the perturbations
increase with the mode number $m$, while it does not depend on the mode
number $n$. Therefore, in a simulation where the instability is excited using Gaussian
noise, we expect the perturbations to grow fastest in the lower region of the
computational domain. This is indeed the case, as we will see in Section \ref{sec:inner_icm}.

\begin{figure}[t]
\centering
\includegraphics{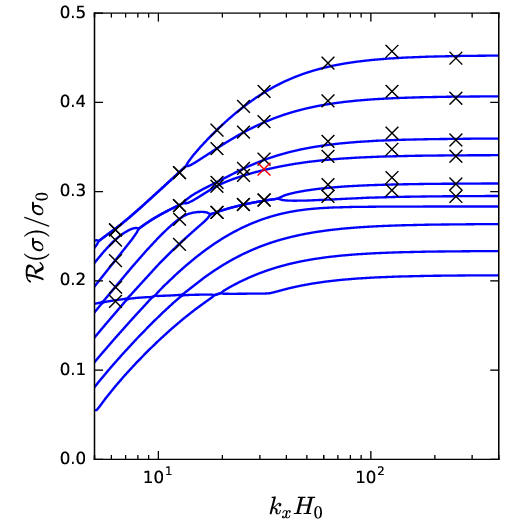}
\caption{
Growth rates as a function of the horizontal wavenumber, $k_x$, for the 10
fastest growing modes. The solid blue lines were obtained using the
pseudo-spectral method. Each cross corresponds to a simulation were the
eigenmodes were used for initial conditions. The numerical growth rate was
found from the subsequent exponential evolution.
}
\label{fig:growth_vs_kx}
\end{figure}

The growth rates as a function of the wavenumber, $k_x$, are shown in Figure
\ref{fig:growth_vs_kx}, where the solid lines are obtained using the
pseudo-spectral method and the crosses are obtained from Athena simulations (see next
subsection). The maximum growth rate is $\sigma_{\rm max} \approx 10 \textrm{
Gyr}^{-1}$ implying that the instability can develop significantly on relevant
time scales. The gradient in the mean molecular weight acts to increase the
growth rate with respect to the homogeneous case. The maximum growth rate
found for the HPBI is however still smaller than the one found for the HBI in
\cite{Lat12} due to the shallower temperature gradient that we use in this
work, which is inspired by the models in  \cite{Pen09} as detailed above. The
temperature ratio, $T_Z/T_0$, is 2.5 in the model of \cite{Lat12} while it is
only $\sim 1.65$ in our model. The growth rate is proportional to the gradient
in temperature in the local, weak magnetic field and fast heat conduction
limit. A rough estimate is therefore that their growth rate should be higher
by a factor of roughly $1.5$, which we confirmed with the results of the full
quasi-global analysis.

\subsection{Solutions obtained with Athena}
\label{subsec:Athena}

We have modified the publicly available MHD code Athena
\citep{stone_athena:_2008} in order to be able to describe the nonlinear
evolution of weakly-collisional atmospheres with non-uniform composition. Our
modified version of the code has anisotropic heat conduction and diffusion of
composition with spatially dependent transport coefficients. These
modifications, along with tests, were described in detail in \cite{Ber16a}
and applied to local settings, where the vertical extent of the simulation
domains considered was small compared to the relevant scale-height.

\begin{figure}[t]
\centering
\includegraphics{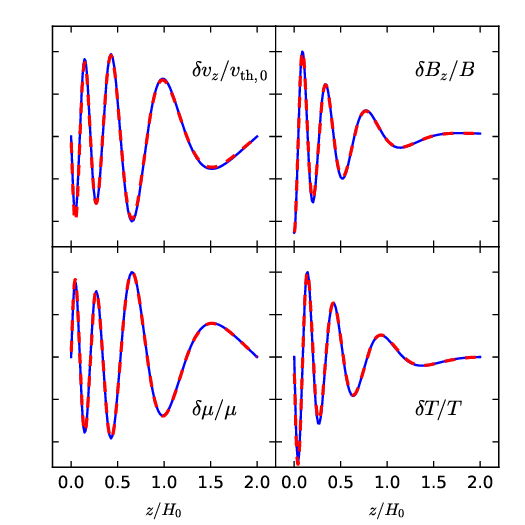}
\caption{
A comparison between a simulation using Athena and the pseudo-spectral method.
A slice in the $x$-direction at $t = 3$ is shown in dashed red and the
pseudo-spectral solution is shown in blue. The growth rate is
$\sigma/\sigma_0= 0.324$
according to the pseudo-spectral method and $\sigma/\sigma_0 = 0.325$
according to the simulation, the error is less than a percent. This mode has
$n = 5$ and $m = 4$ and it is also indicated with a red cross in Figure
\ref{fig:growth_vs_kx}.
}
\label{fig:perturbations}
\end{figure}

We use a pseudo-spectral method in order to test the numerical
solutions obtained with Athena in a quasi-global setting. This is done by
exciting the HPBI in the Athena simulations using an exact eigenmode found in the quasi-global linear
theory. An example is shown in Figure \ref{fig:perturbations} where four of
the components of the perturbation are shown for the $n = 5$ and $m = 4$ mode
at $t = 3$ in units of $t_0 = H_0/v_{\rm 0, th} = 45$ Myr. The growth rate
can be
measured in the simulations by following the evolution of the amplitude of the
perturbations. We find
that the error in the
growth rate for this simulation
is less than a percent compared to the pseudo-spectral linear theory.
We have thoroughly tested this by running a suite of simulations
where we vary the $n$ and $m$ mode numbers. A total of 42 simulations were run
with $n = 1$ to 8 and $m=1$ to 6 with six of the simulations being
degenerate\footnote
{They are complex conjugate solutions so the eigenvalues do differ in their
imaginary part which is however very small.} in $m$. A comparison between the
growth
rates obtained using the pseudo-spectral method and the simulations is shown in
Figure \ref{fig:growth_vs_kx}.
In this figure, each cross corresponds to an individual
simulation where the eigenmode was excited exactly.
The results shown in the figure were obtained by fitting
an exponential function to the time evolution of the volume average of $\delta
B_x^2/8\pi$ for each simulation. We have checked that the other
components of the perturbation (such as $\delta v_x$ and $\delta \mu/\mu$)
also grow at the correct rate. Due to the exact excitation of the
modes the evolution is exponential from the onset.

The simulations were run at half the resolution, $256\times512$, of
the simulations presented in Section
\ref{sec:inner_icm} in order to expedite the numerical simulations.
As noted already by \cite{Lat12}, the low resolution causes some discrepancy at
the highest wavenumbers. These tests illustrate the generally good agreement between the numerical
implementation in Athena and our quasi-global linear theory using pseudo-spectral methods.

\section{Inner regions of the ICM}
\label{sec:inner_icm}

\begin{figure*}[t]
\includegraphics{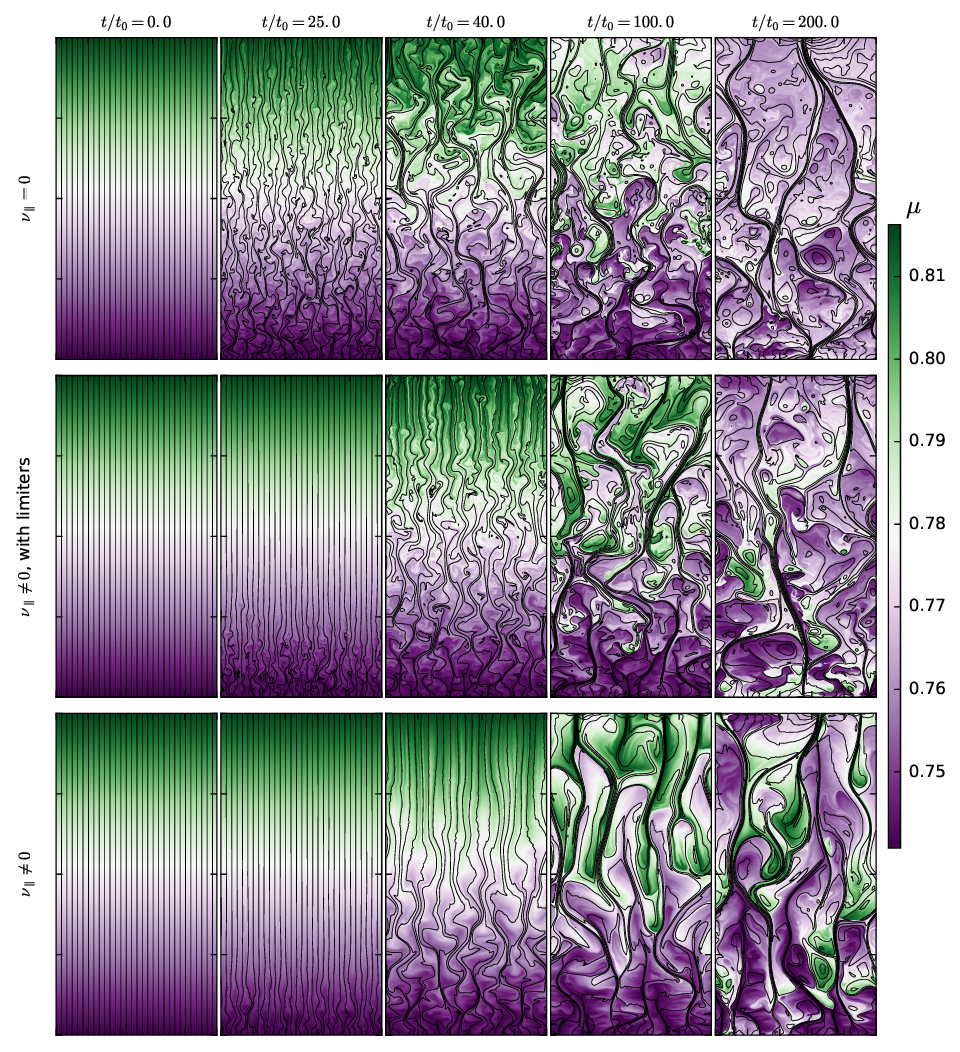}
\centering
\caption{
Evolution of the HPBI as a function of time in units of $t_0=H_0/v_{\rm 0, th} =
45$ Myr. The size of the box is $H_0 \times 2 H_0$ with $H_0 = 40$ kpc.
The bottom of the atmosphere has $T_0 = 5.8$ keV and $c_0 = 0.52$ while
the top of the atmosphere has $T = 9.6$ keV and $c = 0.62$, values found
at $r_0 = 160$ kpc and $r = r_0 + 2H_0 = 240 $ kpc in the model of
\cite{Pen09}. The top row of panels include anisotropic heat conduction
and the middle and bottom rows also include Braginskii viscosity. The
middle row uses limiters. An animated version is available at
\href{http://www.nbi.dk/~berlok/movies/icm_quasi_global.html}{this http URL.}
}
\label{fig:hpbi_icm_nonlinear}
\end{figure*}

In this section we consider the long time evolution of plasma instabilities in
the inner regions of the ICM by performing fully nonlinear simulations of
the HPBI using Athena. We describe the  details of
the numerical setup in Section \ref{subsec:numerical_setup}, the evolution of key
quantities in Section \ref{subsec:evolutions} and conclude by comparing a
simulation of the HPBI with a simulation of the HBI in Section
\ref{subsec:hbi_comparison}.

Some of the simulations include Braginskii viscosity which is a numerical and
theoretical challenge \citep{schekochihin_plasma_2005,Kun12}.
The reason is that microscale instabilities are triggered if the magnitude
of the pressure anisotropy grows too large. The pressure tensor can
become anisotropic due to conservation of the first adiabatic invariant of the
ions. Specifically, when the pressure anisotropy exceeds the limits given by
\be
-\f{B^2}{4\pi} < p_\perp - p_\para < \f{B^2}{8\pi} \ .
\label{eq:fire_and_mirror}
\en
the firehose or mirror instabilities are triggered \citep{1958PhRv..109.1874P}.
This is an issue for the kind of simulations we consider because the
microscale instabilities are not correctly described by Braginskii-MHD
\citep{schekochihin_plasma_2005}.

Two different approaches used to handle this problem are described in \cite{Kun12}.
The first approach is to use high resolution and hope that the firehose
instability grows sufficiently fast in order to regulate the pressure
anisotropy. The second approach is to artificially limit the pressure anisotropy
to the interval given by Equation (\ref{eq:fire_and_mirror}). The latter
approach is motivated by studies showing that the microscale instabilities will
likely saturate by driving the pressure anisotropy back to marginal stability
\citep{2008PhRvL.100h1301S,Bal09,2011MNRAS.413....7R} and a similar approach
has been used for the magnetorotational instability in local studies of weakly
collisional disks \citep{Sha06}. A recent study of the solar wind showed
that the firehose and  mirror thresholds also provide good constraints for the
pressure anisotropy in a multi-species plasma with electrons, hydrogen and
helium ions \citep{2016ApJ...825L..26C}. We have used both of these
approaches
in our simulations.

\subsection{Description of numerical setup and overview of results}
\label{subsec:numerical_setup}

We perform a suite of three simulations of the HPBI: one without Braginskii
viscosity
(HPBI\_isoP), one with Braginskii viscosity where the pressure anisotropy is
limited (HPBI\_Blim) by Equation (\ref{eq:fire_and_mirror}) and one with
Braginskii viscosity without limiters (HPBI\_Brag). We furthermore consider a
simulation where the atmosphere has uniform composition ($c_0 = c_Z = 0.62$)
in order to compare the simulations with the homogeneous case (HBI\_Brag) in
Section \ref{subsec:hbi_comparison}.

\begin{figure*}[t]
\includegraphics{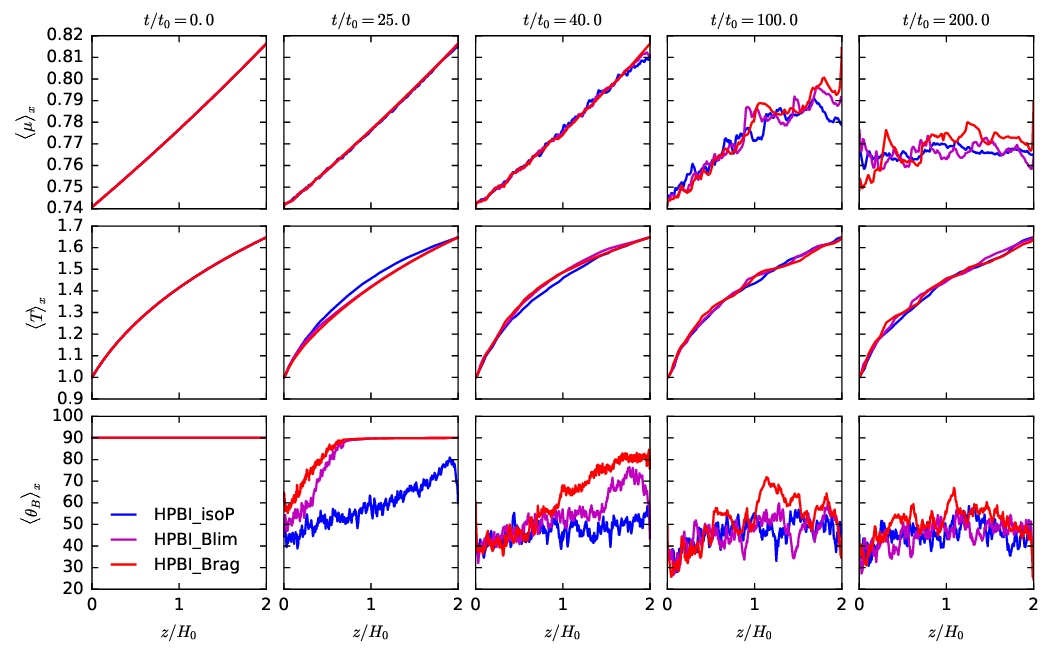}
\centering
\caption{Key quantities averaged along $x$ as a function of height $z$
scaled by $H_0 = 40$ kpc, at times $t/t_0 = 0$, 25, 40, 100
and 200, with $t_0 = 45$\,Myr, for the simulation without viscosity (blue),
with viscosity and limiters (magenta) and with viscosity but without limiters (red).
\emph{Top:} Average mean molecular weight. \emph{Middle:} Average temperature.
 \emph{Bottom:} Average inclination angle of the magnetic field.
 }
\label{fig:x_averages}
\end{figure*}

The initial condition is the plane-parallel atmosphere introduced in
Section \ref{sec:eq_profile}. The instability is triggered by adding Gaussian,
subsonic noise in the velocity components with a magnitude of $10^{-4}$.
This initial condition, together with the imposed perturbations, ensures an
initial evolution where all the quantities evolve exponentially in time.
In reality the agents driving
the ICM dynamics are more complex and involve, for example,
the stirring by mergers in the outskirts of the cluster and outflows from
active galactic nuclei (AGNs) in the cluster core.
All simulations have a
spatial extent of $H_0 \times 2 H_0$ (with $H_0 = 1$ in code units). The
resolution used is $512\times1024$. In terms of the dimensionless units
introduced above, the coefficients
for anisotropic heat diffusivity and Braginskii viscosity are $\kappa_\para =
1.4\times 10^{-2}\ T^{5/2}\mu\rho^{-1}$ and $\nu_\para = 2.6 \times 10^{-4}\
T^{5/2}\rho^{-1}$.

The three simulations of the HPBI are shown in Figure
\ref{fig:hpbi_icm_nonlinear} where the evolution of the mean molecular weight
and the magnetic field lines is followed as a function of time with snapshots
at $t/t_0 = 0$, $25$, $40$, $100$ and $200$, {with $t_0=H_0/v_{\rm 0, th} = 45$\,Myr}.
In this figure, a high (low) concentration of helium is indicated with green (purple) and the
magnetic field lines are shown as solid black lines. The top row of panels was
created from HPBI\_isoP which did not include Braginskii viscosity while the
middle row (HPBI\_Blim) and bottom row (HPBI\_Brag) of panels did include
Braginskii viscosity.

The various panels in  Figure \ref{fig:hpbi_icm_nonlinear}  illustrate how the magnetic field,
which is initially vertical,
becomes rapidly tangled. The temperature profile, which is not shown here does
not vary significantly as time progresses. The composition evolves
rapidly with bubbles of helium sinking down to the center of the core. By the
end of the simulation, the helium content has been very well mixed in
HPBI\_isoP. In HPBI\_Brag blobs of gas with a high helium content have sunk
towards the center of the cluster but they have retained their structure and
have not mixed with their new environment. This lack of mixing can be
understood from the ability of Braginskii viscosity to make the magnetic field
retain a coherent structure over larger distances than one would find for a
simulation with isotropic pressure (see the discussion of the MTI in
\citealt{Kun12}). This feature, along with the fact that the magnetic field is
tied to the gas, suppresses small scale mixing of the helium content. This
implies that the spatial distribution of helium might be more patchy in a
viscous ICM than in a non-viscous ICM.

\subsection{Evolution of composition, temperature, magnetic field inclination, and energy densities}
\label{subsec:evolutions}

The evolution of the composition gradient can be illustrated by taking
averages along the $x$-direction, designated by the brackets $\langle \rangle_x$. This is
shown in the top row of panels in Figure \ref{fig:x_averages}, which have
been produced at the same times as the snapshots shown in Figure \ref{fig:hpbi_icm_nonlinear}.
We see that on very long timescales of the order of $9$\,Gyr ($t/t_0=200$) the instability, on average,
acts to remove the gradient in composition  that originally gave rise to it. Note, however, that the
gradient in composition is rather long-lived as its profile remains rather unaltered
until about $4.5$\,Gyr ($t/t_0=100$).

We have limited our study to a binary mixture of hydrogen and helium,
the latter being the most important element leading to potential biases in X-ray measurements
\citep{Mar07}. Our approach has the advantage that
enrichment from galaxies can be ignored in our simulations because the mass of
helium in the ICM is much greater than the stellar component (see e.g.
\citealt{And10}). This is not the case for heavier metals for which
enrichment becomes important. A
systematic study of mixing in the presence of imposed MHD turbulence
including the injection of pollutants can be found in \cite{2014ApJ...784...94S}
along with a detailed analysis of the mixing process.

The evolution of the temperature gradient is less dramatic,  as illustrated in the middle row of panels in Figure \ref{fig:x_averages}. Changes in the temperature
profile are also modest when the composition is uniform, as seen in, for
instance, \cite{Kun12} and Figure \ref{fig:hbi_icm_nonlinear}.

\begin{figure*}[t]
\includegraphics{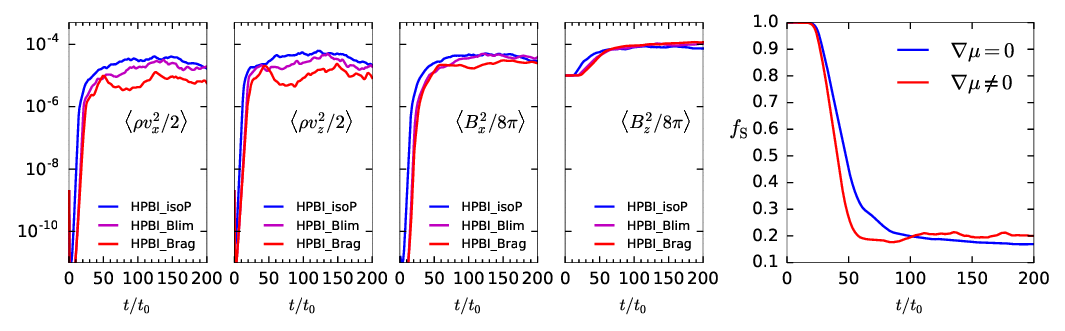}
\centering
\caption{Evolution of the magnetic and kinetic energies in the simulations of
the HPBI with blue (HPBI\_isoP), red (HPBI\_Blim) and magenta (HPBI\_Brag)
with spatial averages of $\rho v_x^2/2$ (first panel), $\rho v_z^2/2$ (second
panel), $B_x^2/8\pi$ (third panel) and $B_z^2/8\pi$ (fourth panel). In the
fifth panel we show the evolution of the volume-averaged Spitzer parameter,
$f_{\rm S} =
\langle Q_c/\tilde{Q}\rangle$ for the HBI and the HPBI. The initial growth of
the HPBI is faster than for the HBI but the final state of the simulation with
a composition gradient has a heat flux that is roughly 20 \% higher than for
the uniform simulation. More information about this panel can be found in
Section \ref{subsec:hbi_comparison}.}
\label{fig:merged_figure}
\end{figure*}
It is already evident from Figure \ref{fig:hpbi_icm_nonlinear} that the
magnetic field changes its inclination with respect to the direction
perpendicular to gravity as time progresses. This can be further studied by
considering the average inclination angle defined by \citep{Par08}
\be
\langle \theta_B \rangle_x = \langle \sin^{-1}(|b_z|) \rangle_x .
\en
The evolution of this quantity is shown in the bottom row of panels in Figure \ref
{fig:x_averages}. We see that the instability grows fastest at the bottom of the
atmosphere but increases its region of influence as time progresses.
The evolution of $\langle \theta_B \rangle_x$ is initially fast compared with the
changes in either composition or temperature and the average angle has changed
significantly at most heights by $1.8$\,Gyr ($t/t_0=40$)\footnote{In the second panel
of the bottom row, it is  evident that Braginskii viscosity slows down the instability.}.
However, on timescales of the order of $4.5$\,Gyr ($t/t_0=100$),
$\langle \theta_B \rangle_x$ seems to settle at around $\sim 30-50^{\circ}$
and the overall distribution of angles shows little evolution until the end of the
run at $9$\,Gyr ($t/t_0=200$).
The change in average inclination angle might have consequences for the cooling flow
problem \citep{Fab94} because heat transport  is primarily along magnetic field
lines. We discuss this in more detail in terms of the Spitzer parameter below.

We consider the evolution of the the kinetic and magnetic energy densities by
calculating volume averages, designated by the brackets $\langle \rangle$, and
shown in Figure \ref{fig:merged_figure} as a function of time. Braginskii viscosity
acts to inhibit the growth rate of the instability and we therefore see the highest
growth rate in HPBI\_isoP, followed by HPBI\_Blim (which is less viscous than
HPBI\_Brag due to the limiters) and then HPBI\_Brag. All the simulations saturate
with magnetic and kinetic energy density components in rough equipartition
with $B_z^2/8\pi$ having increased by a factor of about ten.

\begin{figure*}[t]
\includegraphics{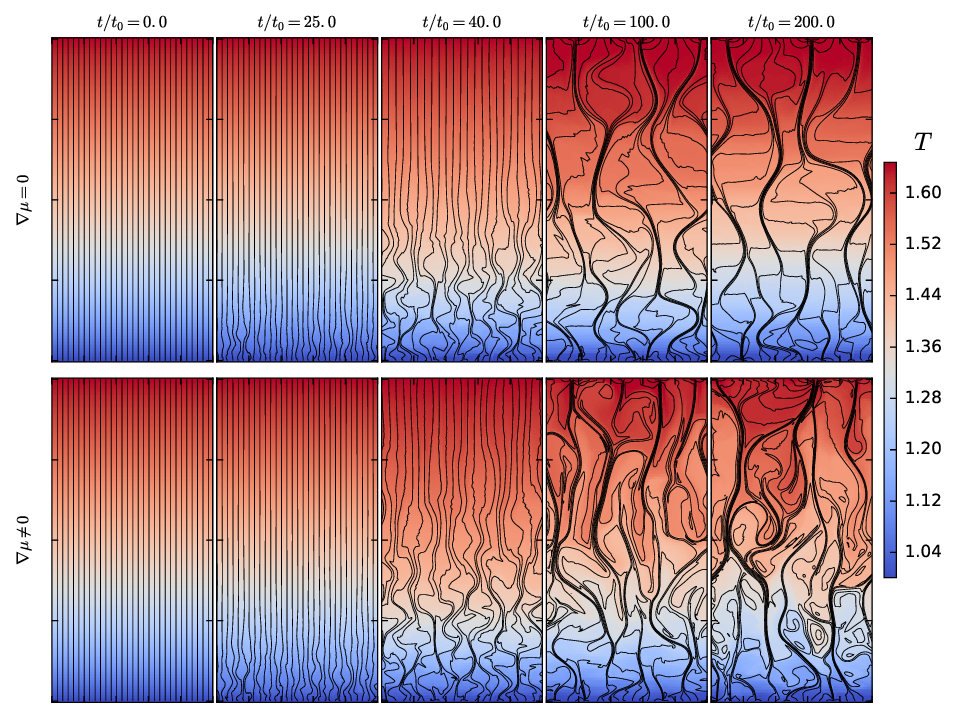}
\centering
\caption{Temperature and magnetic field evolution as function of time for the
HBI (HBI\_Brag, upper row) and the HPBI (HPBI\_Brag, lower row) in units of
$t_0=H_0/v_{\rm 0, th} =
45$ Myr. The size of the box is $H_0 \times 2 H_0$ with $H_0 = 40$ kpc. The
evolution of the composition for HPBI\_Brag is shown in the bottom row of
Figure \ref{fig:hpbi_icm_nonlinear} while HBI\_Brag has uniform composition at
all times. Both simulations included anisotropic heat conduction
and Braginskii viscosity without limiters.}
\label{fig:hbi_icm_nonlinear}
\end{figure*}

\subsection{Comparison between HBI and HPBI}
\label{subsec:hbi_comparison}

In order to compare our results with previous work on the HBI where the plasma
is
assumed to be of uniform composition, we performed a simulation (HBI\_Brag)
using the same atmosphere as for the HPBI but with $c_0 = c_Z$. All other
aspects of this simulation are identical to run HPBI\_Brag. The evolution of the
temperature and the magnetic field is shown in Figure \ref{fig:hbi_icm_nonlinear}.
The gradient in composition leads to a slightly faster growth rate with
respect to the homogeneous case, as predicted by linear theory
\citep{Pes13,Ber15}.

We see that both instabilities reorient the magnetic field, driving the
average inclination of the magnetic field to be more horizontal (azimuthal).
This feature of the HBI has been argued to be of importance for the cooling
flow problem because a magnetic field that is predominantly perpendicular
to gravity tends to insulate the core from heat transport from the
outskirts of the cluster \citep{Par08,Bog09,Par09}. For a vertical (radial)
magnetic field the heat flux is given by the Spitzer value
\be
\tilde{Q} = - \chi_\para\, \D{T}{z} \ ,
\en
while the heat flux is zero for a horizontal (azimuthal) field.
In \cite{Ber16a}, we showed that when considering a local domain in an
isothermal atmosphere, the HPBI drives the magnetic field to have an
average inclination angle of $45^{\circ}$.
This is in contrast to the uniform case, where the HBI drives the magnetic
field towards $0^{\circ}$ \citep{Par08}. This feature of the
HPBI led us to speculate that a composition gradient might be able to
alleviate the cooling flow problem by limiting the average magnetic
field inclination. We can now test this idea in our quasi-global simulations
where the temperature gradient has been obtained from \cite{Vik06}.

We consider the evolution of the volume-averaged Spitzer parameter \citep{Par08}
\be
f_{\rm S} = \langle  \hat{\bb{z}} \bcdot \bb{Q}_{\rm c}/\tilde{Q} \rangle \ ,
\en
as a function of time. This quantity shows how effective heat conduction
is
at transporting heat in the radial direction in the cluster compared to the
case of a vertical (radial) magnetic field (or unmagnetised heat conduction). The same
parameter was used to parameterize the effectiveness of helium sedimentation
in the model of \cite{Pen09}. We show the evolution of $f_{\rm S}$ in
the fifth panel of Figure
\ref{fig:merged_figure} with the HBI simulation ($\nabla \mu = 0$) indicated with
a blue solid line and the HPBI simulation ($\nabla \mu \neq 0$) indicated with
a red solid line. Initially, the volume-averaged Spitzer parameter decreases more rapidly in
the HBPI simulation than in the HBI simulation, the reason being the slower
growth rate in the absence of a composition gradient. The inclusion of a
composition gradient however leads to an increase in the volume-averaged
Spitzer parameter at
late times compared to the homogeneous case. We find that $f_{\rm S} \approx
0.20$ ($f_{\rm S} \approx 0.17$) for the HPBI (HBI) at $t/t_0 = 200$. This
corresponds to an increase in heat flux by $\sim 20 \%$ with respect to the
homogeneous case.

\section{Outer regions of the ICM}
\label{sec:outer_icm}
\begin{figure*}[t]
\includegraphics{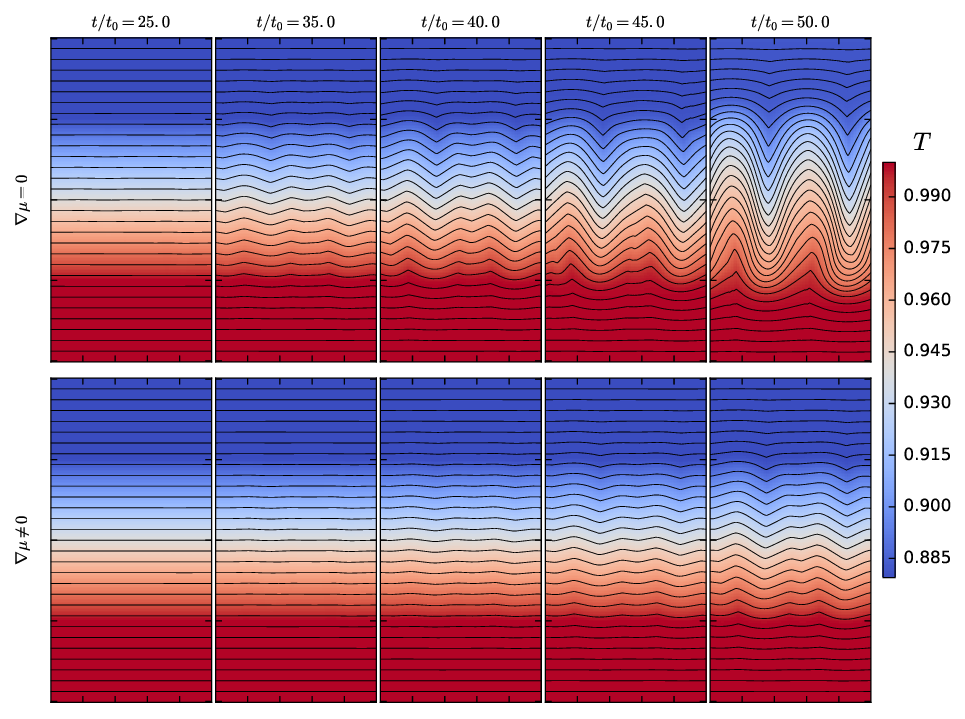}
\centering
\caption{
Evolution of the MTI (upper row) and the MTCI (lower row) as function of time
in units of $t_0=H_0/v_{\rm 0, th} = 230$\,Myr. The size of the unstable part of
the box is $H_0 \times H_0$ with $H_0 = 300$ Mpc. The bottom of the unstable
region has $T_0 = 10.2$ keV while the top of the unstable region has $T = 9$
keV, values found at $r_0 = 0.65$ Mpc and $r = r_0 + H_0 = 0.95$ Mpc in the
model of \cite{Pen09}. The MTI (upper row) has a uniform composition while the
MTCI (lower row) has $c_0 = 0.32$ at the bottom and $c = 0.27$ at the top.}
\label{fig:mtci_icm_nonlinear}
\end{figure*}

In this section, we consider two simulations of the outer parts of the ICM
where we assume that the initial magnetic field is perpendicular to gravity,
i.e., horizontal. In one of these simulations, labeled MTI\_Brag, the plasma
is assumed to have uniform composition and in the other one, labeled
MTCI\_Brag, the composition is assumed to decrease with height (radius).
The bottom of the model plane-parallel atmosphere is located at $r = 0.65$ Mpc, corresponding roughly
to the radius indicated with an \emph {A} in Figure 8 in \cite{Ber15}.

As in Section \ref{sec:inner_icm}, we use reflecting boundary conditions at the top
and bottom of the domain. The MTCI creates vertical motions that will
eventually reach the top and bottom of the domain, potentially
influencing the nonlinear evolution of the instability. In order to circumvent
this problem, we implement a procedure that has successfully been applied to the MTI by
\cite{Par07} and \cite{Kun12}.
We sandwich the unstable layer between two buffer zones in which
we add isotropic heat conduction and viscosity in order to damp any motions making
their way into these regions.  This results in two stable layers at the top and
bottom where
the density decreases exponentially away from the mid-plane and
the temperature and composition are constant, with values
$T_0$ and $c_0$ ($T_Z$ and $c_Z$) in the lower (upper) stable layer.

We use the equilibrium for an initially horizontal magnetic field given in
\cite{Ber16a}, i.e., the temperature and composition profiles given by
\be
T(z) &=& T_{0}+ s_{\rm T}\,z' \ , \\
c(z) &=& c_{0}+ s_{\rm c}\,z' \ ,
\en
in the middle of the computational domain ($z' = z-H_0/4$ and $H_0/4 < z <
3H_0/4$) where $s_
{\rm T} = (T_{\rm Z} - T_0)/H_0$ is the slope in temperature and $s_{\rm
c} = (c_{\rm Z} - c_0)/H_0$ is the slope in composition.
The pressure in the unstable region is given by
\be
P(z') = P_0 \left( \f{T(z')\,\mu(z')}{T_0\,\mu_0}\right)^{\alpha} \ ,
\en
where $\mu(z)$ is related to $c(z)$ by Equation
(\ref{eq:relation_from_c_to_mu}) and the constant coefficient $\alpha$ is given by
\be
\alpha = - \f{T_0}{H_0} \left(\f{4}{4s_{\rm T} + 5 \mu_0 T_0 s_{\rm c}}\right) \ .
\en
Due to the very strong time-step constraint arising from the non-ideal terms,
we have however opted to use significantly reduced values of $\chi_\para$ and
$\nu_\para$, by dividing the expected values by a factor of ten. The dimensionless
values used in the simulations are therefore $\kappa_\para = 0.12\,T^{5/2}\mu
\rho^{-1}$ and $\nu_\para = 3.3 \times 10^{-3} T^{5/2}\rho^{-1}$.
Both temperature and composition decrease with radius with $T_0 = 10.2$ keV
and $c_0 = 0.32$ at the bottom of the unstable domain and  $T = 9$ keV and $c
= 0.27$ at the top for the non-uniform simulation ($c_0=c_Z=0.32$ for the
uniform simulation). Furthermore, we use $\beta = 10^{5}$ and include
Braginskii viscosity without limiters. As for the inner region, these
simulations start from
hydrostatic equilibrium with a subsonic velocity perturbation. In real
clusters, the turbulence caused by accretion of material onto the cluster can
contribute with a significant fraction of the pressure support needed to
counteract gravity \citep{2009ApJ...705.1129L,2014ApJ...792...25N}.

The evolution of the initial
horizontal magnetic field and the temperature is shown in Figure
\ref{fig:mtci_icm_nonlinear} for MTI\_Brag (top row) and MTCI\_Brag (bottom
row). In these simulations the unit of time is $t_0 = H_0/v_{\rm 0, th} = 230$
Myr. Note that because this is roughly a factor of 5 longer than the $t_0$
characterising the inner cluster it is only necessary to evolve the simulations
up to $50 t_0$ to cover timescales of the order of $10+$\,Gyr.
For the MTI (top row) the evolution is very similar to the one
presented for the MTI with Braginskii viscosity in Figure 17 in \cite{Kun12}
except that the growth rate is significantly slower in the simulation
presented here. This is simply due to the much shallower temperature gradient
that we are using \citep{Vik06}. As predicted by theory \citep{Pes13,Ber15}, we observe that
the instability grows at an even slower rate when a gradient in composition is
included (bottom row of Figure \ref{fig:mtci_icm_nonlinear}). For these
parameters, linear theory predicts that the maximum growth rate\footnote
{Found by solving the dispersion relation for a grid of values in $k$-space
and taking the maximum value.} of the MTI is
$\sigma = 1.42$ $\mathrm{Gyr}^{-1}$ while the MTCI only has a maximum growth
rate of $\sigma = 1.09$ $\mathrm{Gyr}^{-1}$.
The maximum growth rate is only $\sim 20 \%$ slower but this difference is
able to significantly alter the final state of the system in this case. The
instabilities are still in the exponential phase and have not reached
saturation at the end of the simulation.
%
\section{Discussion}
\label{sec: Discussion}
Understanding the distribution of helium in the intracluster medium is an
open problem with important implications for astrophysics and cosmology
\citep{Mar07,Pen09}.

The assumption of a spatially uniform composition of helium in the ICM is
routinely applied when interpreting X-ray observations of galaxy clusters.
This can lead to biases in the estimates of various key cluster  parameters if
a composition gradient is present, which can in turn propagate into estimates
of the inferred cosmological constants \citep{Mar07,Pen09}.

Current models to address this problem are one-dimensional and treat the
turbulent, magnetized nature of the medium in a very crude way
\citep{Fab77,Gil84,Chu03,Chu04,Pen09,Sht10}.
This usually amounts to parametrize the presence of magnetic fields so that
its main effect is to slow down the sedimentation process at the same rate at
all radii. The advantage of these kind of models is that
they can be evolved for long timescales. In general terms they predict helium
profiles that peak off-centre when the (fixed) temperature profile is typical
of cool-core clusters \citep{Pen09}.

The approach employed, however, does not allow to take into consideration the
fact that the weakly-collisional nature of the ICM renders its properties
anisotropic due to inefficient transport across the magnetic field. When the
effects of anisotropic heat conduction, viscosity and particle diffusion are
considered, with given initial thermal and composition profiles, a wide
variety of instabilities, which are absent in one-dimensional settings, can
play an important role in the plasma dynamics
\citep{Bal00,Bal01,Qua08,Kun11,Pes13}.

Ideally, it would be desirable to evolve, in a global setting, the system of
equations that describes the evolution of the thermal and composition
gradients of a weakly-collisional plasma (with initial cosmic composition) in
the potential well of a dark-matter halo. In lieu of pursuing this arguably
daunting task at once, we have opted to analyze this problem by
developing a better understanding of the plasma dynamics using a number of
approximations, which can in principle be relaxed in future studies:

\emph{i)} We have adopted as a working model a binary mixture composed of
hydrogen and helium in the Braginskii-MHD approximation with Braginskii
viscosity \citep{Bra,Pes13}. This approach is known to be subject to
small-scale
instabilities
that need to be dealt with appropriately in numerical simulations
\citep{schekochihin_plasma_2005,Kun12}. The firehose and mirror
instabilities
can be more accurately captured using hybrid particle-in-cell codes where the
ions are treated as particles and the electrons are treated as a fluid (see
for instance \citealt{2014PhRvL.112t5003K}). Studies of the firehose and
mirror instabilities normally
assume a hydrogen plasma. Extending such simulations in order to study the
kinetics of a multispecies plasma might therefore give new insights on how to
incorporate microscale instabilities in Braginskii-MHD simulations of binary
mixtures.

Particle-in-cell simulations have shown that the heat
conductivitity, $\chi_\para$, can be significantly reduced by the action of
the ion mirror instability \citep{2016MNRAS.460..467K,2016ApJ...824..123R},
the latter study also finding a reduction due to the electron whistler
instability. The suppression of conductivity is found by
\cite{2016MNRAS.460..467K} to be due to a combination of trapping of electrons
in $\delta B/B\sim 1$ magnetic traps and a decreased mean free path of
collisions due to pitch-angle scattering off microscale fluctuations. The
suppression of heat flux with respect to the Spitzer value in the simulations
we present in Section \ref{subsec:hbi_comparison} is due to the
change in magnetic field orientation. Both of these effects yield a
suppression of the heat flux by a factor of $\sim 5$, a result which has also
been found in magnetic turbulence
\citep{2001ApJ...562L.129N,2004ApJ...602..170C}. If a large fraction of the
plasma is mirror unstable, the two effects could in principle act in unison to
give a total reduction by a factor of $\sim 25$. A reduction of $\chi_\para$
due to microscale instabilities could however also have consequences for the
importance of the HBPI and the MTCI, given that their growth rates depend
on fast heat conduction along magnetic field lines.

More recently,~\cite{2016arXiv160805316X} studied the stability of
a collisionless, thermally stratified plasma by using linear Vlasov theory
and described the kinetic version of the MTI. Moreover, an electron version
of the MTI, the eMTI, was found to operate at sub-ion-Larmor scales and have a
faster growth rate than the long wavelength kinetic MTI.
The dispersion relation
derived in~\cite{2016arXiv160805316X} can in principle be used to determine
whether the MTCI also carries over to the collisionless regime. Subsequent
particle-in-cell simulations could then be used to assess the differences with
respect to the Braginskii-MHD framework employed in this paper. At present,
such comparisons are yet to be done in homogeneous settings. The nonlinear
outcome of the combined presence of the long wavelength kinetic MTI and the
eMTI has yet to be explored by dedicated particle-in-cell in simulations.

\emph{ii)} We have simplified the geometry of the problem by considering a
plane-parallel atmosphere. This approach has been applied with success in a
wide variety of astrophysical settings. Its accuracy depends on the phenomena
under study having radial scales that are smaller than the fiducial radius at
which the model is adopted. We have improved on our previous work
\citep{Pes13,Ber15,Ber16a}, in which the domain under considerations were
local in both radius and azimuth, and developed a quasi-global approach,
extending previous work in homogeneous settings \citep{Lat12,Kun12}.
This enabled us to consider domains that are not necessarily
small compared to the thermal scale height at the fiducial radius.

\emph{iii)} The helium concentration profile in galaxy clusters is
unfortunately not directly observable \citep{Mar07}.
In order to construct our model
atmospheres we relied on current one-dimensional helium sedimentation models
\citep{Pen09}.
By considering these as initial conditions, we investigated the evolution of a
number of instabilities that feed off the gradients in temperature and
composition in the inner regions as well as the outskirts of the ICM.
This approach assumes that the timescales for the evolution of the temperature
and composition profiles are long compared to the timescales for the
instabilities to grow significantly \citep{Ber15}.
While this seems to be a reasonable
assumption, the fact that the large scale gradients, from which the
instabilities feed off, are unable to evolve prevents us from understanding
how the instabilities interact with the processes that drive their evolution
at a more fundamental level \citep{1969fecg.book.....B,1990ApJ...360..267B}.

While our approach cannot directly predict the evolution of the helium distribution
in the ICM, we have been able to learn a few interesting things about how composition
gradients can influence the dynamics of the weakly-collisional medium.

In the inner region of the ICM, the nonlinear evolution of our model
shows that helium rich material, initially at the top of the atmosphere,
will fall down onto the inner core of the cluster. The relevant timescale
for mixing to occur is of the order of a few Gyrs. It is important to emphasise
that this process cannot be attributed to standard convection driven by
composition gradients since the Ledoux criterion \citep{1947ApJ...105..305L}
is fulfilled in the model atmospheres we employed. The driving mechanism is
the generalisation of the  heat-flux driven buoyancy instability HBI, which we
have termed the heat- and particle-flux-driven buoyancy instability (HPBI).
We analysed in some detail the effects that the evolution of the magnetic fields
has on the thermal conductivity of the plasma to assess whether composition gradients
can alleviate the core insulation observed in homogeneous settings \citep
{Par08}.
Beyond a few
Gyrs, the average inclination angle of the magnetic field is close to
$\sim\,30-50^\circ$
resulting in an averaged Spitzer parameter higher by about 20\% than the value
obtained in a corresponding homogeneous simulation. The distribution of
composition is more patchy in the simulations where Braginskii viscosity
is included because it can inhibit mixing at smaller scales. The main
conclusions described here however seem to be rather insensitive to Braginskii
viscosity when composition is averaged along the azimuthal direction.

We also investigated the dynamics of the outskirts of the
ICM, where both the temperature and the composition are expected to decrease
with increasing radius \citep{Vik06,Pen09}. In this case, the mechanism driving
instabilities is
the generalisation of the  magneto-thermal instability (MTI), which we have
termed the magneto-thermal-compositional instability (MTCI).  The shallower
gradients characterising current models imply that the instabilities evolve
slowly and there is not enough time for them to evolve in the strong
non-linear regime even after several Gyrs. Therefore, in the outskirts of the
cluster, the instabilities are rather inefficient at erasing the composition
gradients.

This mismatch between fast growing instabilities in the inner core and rather
slow instabilities in the outskirts could imply that compositions gradients in
cluster cores might be shallower than predicted by one-dimensional models. One
could speculate about the long-term outcome of the interplay between the
various competing processes, but it seems to be safer to develop more
self-consistent models in which the instabilities can develop in a global setting
where the physics driving helium sedimentation is accounted for. One
alternative, intermediate step in developing these models could consist of
using the  type of numerical simulations we have employed here to develop more
physically motivated effective models for mixing that can be incorporated in
improved one-dimensional models.

The weakly-collisional, magnetized nature of the ICM is likely to have an
impact on the long-term evolution of the gas dynamics, including the issue of
whether helium can sediment efficiently. Our work constitutes the first few
steps in this direction. More quantitative statements will demand improved
models that incorporate the physics driving the sedimentation process while
simultaneously accounting for the anisotropic transport properties of the
medium.

\acknowledgements
We acknowledge useful discussions with Daisuke Nagai, Matthew Kunz, Prateek
Sharma, Ellen Zweibel, and Ian Parrish during the \emph{3\textsuperscript{rd}
ICM Theory and Computation Workshop} held at the Niels Bohr Institute in 2014.
We would like to thank Matthew Kunz for sharing with us the Athena files used
to run the simulations presented in \cite {Kun12}.
Furthermore, we thank him and Henrik Latter for providing us with a copy of
the script they used in \cite{Lat12} and that we used as a starting
point to build the code used in this paper.
We
are grateful to Daisuke Nagai and Fang Peng for sharing with us the Fortran
program used to calculate the mean molecular weight profile used as
inspiration for the simulations presented in this paper. We also
acknowledge useful discussions with Gopakumar Mohandas and Sagar
Chakraborty.
We thank the anonymous referee for helpful comments that helped improve the final version of the manuscript. The research leading
to these results has received funding from the European Research Council under
the European Union's Seventh Framework Programme (FP/2007-2013) under ERC
grant agreement 306614. T.B. also acknowledges support provided by a L{\o}rup
Scholar Stipend and M.E.P. also acknowledges support from the Young
Investigator Programme of the Villum Foundation.

\bibliography{Bibliography}

\end{document}